\documentclass[journal]{IEEEtran}

\usepackage{amsmath}

\usepackage[ruled]{algorithm2e}

\usepackage{amsmath,amssymb,amsfonts}
\usepackage{algorithmic}
\usepackage{array,color}
\usepackage{bm}
\usepackage{float}
\usepackage{threeparttable}
\usepackage{color}
\usepackage{bbding}
\usepackage{subfig}
\usepackage[font=small,labelsep=period]{caption}
\usepackage{array,color}
\usepackage{colortbl}
\usepackage{amsfonts,amssymb}
\usepackage{CJK}
\usepackage{booktabs}
\usepackage{multirow}
\usepackage{url}
\usepackage{stfloats}
\usepackage{diagbox}
\usepackage{amsmath}
\usepackage{autobreak}
\usepackage{cite}
\usepackage{todo}

\usepackage{float}

\ifCLASSINFOpdf
\usepackage[pdftex]{graphicx}
\graphicspath{{../pdf/}{../jpeg/}}
\DeclareGraphicsExtensions{.pdf,.jpeg,.png}
\else
\usepackage[dvips]{graphicx}
\graphicspath{{../eps/}}
\DeclareGraphicsExtensions{.eps}
\fi
\begin{document}

\title{User Pairing and Power Allocation for FTN-based SC-NOMA and MIMO-NOMA Systems Considering User Fairness}

\author{Peiyang~Song}

\markboth{Journal of \LaTeX\ Class Files,~Vol.~14, No.~8, August~2015}%
{Shell \MakeLowercase{\textit{et al.}}: User Pairing and Power Allocation for MIMO-NOMA considering user fairness}

\maketitle

\begin{abstract}
	The development of the industrial Internet of Things (IoT) calls for higher spectrum efficiency (SE). Faster than Nyquist (FTN) and non-orthogonal multiple access (NOMA) are both promising paradigms to improve the SE without any extra spectrum resources required. The combination of FTN and NOMA is an interesting issue and has been focused on recently. In the NOMA technology, user pairing and power allocation are key algorithms determining system capacity. This paper first proposes a joint user pairing and power allocation algorithm for the FTN-based single-carrier (SC) NOMA system. Then, the FTN-based multiple-input-multiple-output (MIMO) NOMA is studied and a dynamic user pairing and power allocation scheme is presented. In both scenarios, the maximum available sum rate (ASR) is the target. While based on the fairness principle, the user's SE in the NOMA system is guaranteed to be no less than that in the OMA system. Simulation results show the advantage of the FTN-based NOMA with the proposed scheme in ASR and quality of service (QoS) performance. As far as we know, this paper is the first solution to the issue of user pairing and power allocation in FTN-based NOMA, which proves the great advantage of the combination of these two state-of-the-art technologies.
\end{abstract}


\IEEEpeerreviewmaketitle

\section{Introduction}

\IEEEPARstart{T}{he} past few decades have witnessed the exponential increase of data traffic and the rapid development of communication technology. Nowadays, the spectrum resources are becoming more and more valuable. With the rise in the number of sensors and the variety of transmitted data, the industrial Internet of Things (IoT) requires a higher transmission rate to carry more business and data. In this context, non-orthogonal multiple access (NOMA), working as a powerful tool to improve the spectrum efficiency (SE), attracts a lot of attention from the industrial fields \cite{liu2019noma, noh2021delay, qian2020noma} and has been a promising technology in the fifth-generation (5G) \cite{ding2017survey, cai2017modulation, yuan20205g} and the sixth-generation (6G) communications \cite{liu2022evolution}.

Different from conventional orthogonal multiple access (OMA) systems, NOMA can allocate one frequency bandwidth to several users simultaneously, which requires effective user pairing and power allocation algorithms \cite{masaracchia2019pso, wang2020location, yu2019multiuser}. For example, \cite{masaracchia2019pso} recently proposes a particle swarm optimization (PSO)-based approach to minimize the transmitted power while guaranteeing the quality of service (QoS) of each user. Authors in \cite{wang2020location} develop a low-complexity user pairing scheme based on the location of different users. The impact of the sub-optimal detector in the receiver is studied in \cite{yu2019multiuser} and two corresponding user pairing schemes are proposed to minimize the pair-wise error probability (PEP). 

Also, motivated by its high available sum rate (ASR), NOMA is applied in many other scenarios, such as edge computing \cite{noh2021delay}, intelligent reflecting surface (RIS) \cite{zheng2020intelligent}, full-duplex (FD) communications \cite{liang2020user} and so on. NOMA has been a promising paradigm for improving the capacity of future communication systems.

Faster than Nyquist (FTN) signaling was first proposed by Mazo in the 1970s, which can improve the SE without any spectrum resources required. In recent years, it has been rediscovered and focused on by both the academic and industrial fields \cite{fan2017faster, ishihara2021evolution, song2019blind}. Compared to conventional Nyquist-criterion transmission, FTN signaling transmits symbols at a higher rate and introduces unavoidable inter-symbol interference (ISI) that can be eliminated by proper detection algorithms \cite{bedeer2017low, sugiura2013frequency, song2020dl}.

Inspired by the high SE of both FTN and NOMA, the combination of the two technologies is quite an interesting issue to improve the system ASR further. Although \cite{yuan2020iterative} has pointed out the advantage of the FTN-based NOMA from the theoretical perspective, designing an effective scheme to achieve its performance gain is still an open challenge. For example, a new problem is that the integral operation in the calculation of system capacity makes the FTN-NOMA  different from the conventional Nyquist-NOMA when designing resource allocation algorithms. This paper proposes two user pairing and power allocation algorithms for the FTN-based single carrier (SC) NOMA and multiple-input-multiple-output (MIMO) NOMA, respectively, considering user fairness. And as far as we know, this is the first available solution to the user pairing and power allocation issue for the FTN-based NOMA.

The contribution of this paper is summarized as follows.

\begin{itemize}
	\item We derive an effective power allocation scheme for the 2-user FTN-based SC-NOMA system to maximize the ASR while considering user fairness.
	\item By generalizing the 2-user system to 4-user and 2K-user scenarios, we derive the optimal user pairing scheme to obtain the highest ASR for the FTN-based SC-NOMA.
	\item A dynamic user pairing scheme with effective power allocation for the FTN-based MIMO-NOMA is proposed, considering user fairness.
	\item We conduct comprehensive simulations to evaluate the capacity and QoS performance of our proposed schemes.
\end{itemize}

The rest of this paper is organized as follows. Section \ref{sec:sc_model} introduces the system model of FTN-based SC-NOMA. Then, in Section \ref{sec:up_sc_noma}, a power allocation and corresponding optimal user pairing scheme for FTN-based SC-NOMA is proposed. The model of FTN-based MIMO-NOMA is presented in Section \ref{sec:mimo_model}. And then Section \ref{sec:up_mimo_noma} proposes a dynamic user pairing and power allocation scheme for the FTN-based MIMO-NOMA. Comprehensive simulations on the proposed schemes are demonstrated in Section \ref{sec:simulation}. And finally, Section \ref{sec:conclusion} concludes this paper.


\section{System Model of FTN-based SC-NOMA} \label{sec:sc_model}
This section considers the FTN-based downlink SC-NOMA with $2K$ users uniformly deployed in a disc $\cal D$ with radius $d$. Without loss of generality, we assume that the base station (BS), located at the center of $\cal D$, knows exactly each user's channel state information (CSI). To improve the SE, NOMA requires the $2K$ users to be divided into $K$ groups to share the same resources within each group.

For the user pair containing User-$m$ and User-$n$ ($1\le m\le n\le2K$), the $i$-th signal transmitted by the BS can be written as
\begin{equation}
s_i^{(m,n)}=\sqrt{\epsilon _{m}}s_{i,m}+\sqrt{\epsilon _{n}}s_{i,n},
\end{equation}
where $s_{i,k}$ denotes the $i$-th signal for User-$k$. $\epsilon_m$ and $\epsilon_n$ are power allocation coefficients for User-$m$ and User-$n$ respectively and $\epsilon_m+\epsilon_n=1$.

Compared to conventional Nyquist-NOMA, the FTN-NOMA accelerates the transmission at the price of unavoidable ISI. The $i$-th received symbol for User-$k$ ($k=m,n$) can be written as

\begin{equation}
	\begin{aligned}
		r_{i,k}^{(m,n)}&=\sum_{j=-\infty}^{+\infty}{h_k s_{i-j}^{(m,n)}}g(\left( i-j \right) \alpha T_N)+\tilde{n}_{i,k}\left( i\alpha T_N \right)
		\\
		&=h_ks_i^{(m,n)} g(i\alpha T_N) +\tilde{n}_{i,k}\left( i\alpha T_N \right) \\
		& \,\,\,\,\,\,\,\, +\sum_{j=-\infty ,j\ne 0}^{+\infty}{h_ks_{i-j}^{(m,n)}}g\left( \left( i-j \right) \alpha T_N \right),
	\end{aligned}
\end{equation}
where $h_k=\tilde h/\sqrt{(d_k)^3}$ is the channel gain for User-$k$. $\tilde h$ obeys the Rayleigh distribution and $d_k$ is the distance between User-$k$ and the BS. $0<\alpha\le 1$ is the symbol packing ratio \cite{fan2017faster} in FTN signaling which determines the transmission rate and the strength of ISI. $T_N$ denotes the Nyquist symbol period which can be obtained by $T_N=1/(2W)=(1+\beta)/(2W_T)$, where $W$ and $W_T$ are the available bandwidth and the system's total bandwidth, respectively. $\beta$ is the roll-off factor of the shaping and matched filter. $g(t)$ is obtained by the convolution of shaping and matched filter. $\tilde n_{i,k}(t)$ is the colored noise obtained by the convolution between white noise and the matched filter function. And the power spectral density (PSD) of the white noise is $N_0$.

To simplify the analysis, among $2K$ users, we assume that $|h_1|\le|h_2|\le\cdots\le|h_{2K}|$. User-$m$ is the first user (weak user) in the $(m,n)$-pair while User-$n$ is the second user (strong user). When User-$n$ receives the signal from the BS, the information for User-$m$ will be decoded and removed first to eliminate the interference. Based on the principle of FTN \cite{rusek2009constrained} and NOMA, the rate of User-$n$ can be written as

\begin{equation} \label{eq:sc_rate_near}
	\mathrm{R}_{n}^{\left( m,n \right)}=\frac{1}{\pi \alpha(1+\beta)}\int_0^{\pi}{\log _2\left( 1+\epsilon _n\gamma \left| h_n \right|^2G\left( \omega \right) \right) \rm{d}\omega},
\end{equation}
where $\gamma=2P/N_0$ and $P$ is the power allocated to $(m,n)$-pair.
$G(\omega)\in\left[0, T_N\right]$ is the frequency domain sampled filter function which can be expressed as

\begin{equation}
	G(\omega)= \sum_{k=-\infty}^{\infty}\left|\tilde G\left(\frac{\omega}{2 \pi \alpha T_N}+\frac{k}{\alpha T_N}\right)\right|^{2},
\end{equation}
where $\tilde G(\omega)$ represents the Fourier transform of the shaping filter function.

Especially, as presented by (20) in \cite{rusek2009constrained}, when Nyquist transmission is considered, \eqref{eq:sc_rate_near} turns to be
\begin{equation}
	\mathrm{R}_{n, \rm Nyquist}^{\left( m,n \right)}=\frac{1}{ (1+\beta)}\cdot{\log _2\left( 1+\epsilon _n\gamma \left| h_n \right|^2T_N \right)} \le R^{(m,n)}_n
\end{equation}

Meanwhile, User-$m$ directly decodes its own signal and the signal for User-$n$ is considered as the interference. The available rate of User-$m$ can be written as
\begin{equation} \label{eq:sc_rate_far}
	\begin{aligned}
	\mathrm{R}_{m}^{\left( m,n \right)}=&\frac{1}{\pi \alpha(1+\beta)} \cdot \\
	 &\int_0^{\pi}{\log _2\left( 1+\frac{\left| h_m \right|^2\left( 1-\epsilon _n \right) G\left( \omega \right)}{\epsilon _n\left| h_m \right|^2G\left( \omega \right) +\gamma ^{-1}} \right) \rm{d}\omega}.
	\end{aligned}
\end{equation}

Finally, considering the multiplexing loss $1/2$, the available rates for user-$k$ ($k=1,2\cdots,2K$) in the OMA system can be obtained as
\begin{equation} \label{eq:sc_oma_cap}
	\mathrm{R}_{k}^{OMA}=\frac{1}{2\pi \alpha (1+\beta)}\int_0^{\pi}{\log _2\left( 1+\gamma \left| h_k \right|^2G\left( \omega \right) \right) \rm{d}\omega}.
\end{equation}

Also, when Nyquist transmission is considered, \eqref{eq:sc_oma_cap} turns to be
\begin{equation}
	\mathrm{R}_{k, \rm Nyquist}^{OMA}=\frac{1}{2 (1+\beta)}\cdot{\log _2\left( 1+\gamma \left| h_k \right|^2T_N \right) }.
\end{equation}

\section{User Pairing and Power Allocation for FTN-based SC-NOMA} \label{sec:up_sc_noma}
This section firstly considers the power allocation problem in the 2-user FTN-based SC-NOMA system and derives an available power allocation coefficient to maximize the ASR and guarantee user fairness. Then, the scenario is generalized to 4-user and 2K-user FTN-based SC-NOMA to obtain the optimal user pairing scheme with the derived power allocation within each pair.

\subsection{Power Allocation for 2-user FTN-based SC-NOMA System} \label{sub:up_2_user}
Here, we consider an FTN-based SC-NOMA with only 2 users named User-$m$ and User-$n$. The target of power allocation is to maximize the ASR of two users while guaranteeing the rate of each user is not less than that in the OMA system. 

To meet the rate requirements for User-$m$, the range of $\epsilon_n=1-\epsilon_m$ should satisfy
\begin{equation} \label{eq:tuilun_1}
	\begin{aligned}
		&\mathrm{R}_{m}^{\left( m,n \right)}\ge \mathrm{R}_{m}^{OMA}
		\\
		&\Leftrightarrow  \int_0^{\pi}{\log _2\left( \frac{\sqrt{1+\,\gamma \,\left| h_m \right|^2G\left( \omega \right)}}{\,1+\varepsilon _n\,\gamma \,\left| h_m \right|^2G\left( \omega \right)} \right)}\mathrm{d}\omega \ge 0
		\\
		&\Leftarrow  \frac{\sqrt{\,1+\gamma \,\left| h_m \right|^2G\left( \omega \right)}}{1+\,\varepsilon _n\,\gamma \,\left| h_m \right|^2G\left( \omega \right)}\ge 1  \left( 0\le \omega \le \pi \right) 
		\\
		&\Leftrightarrow  \epsilon _n\le \frac{\sqrt{1+\,\gamma \,\left| h_m \right|^2G\left( \omega \right)}-1}{\gamma \,\left| h_m \right|^2G\left( \omega \right)}\,\, \left( 0\le \omega \le \pi \right).
	\end{aligned}
\end{equation}

Here, we set $f\left( x \right) =\frac{\sqrt{1+\,\gamma \,\left| h_n \right|^2x}-1}{\,\gamma \,\left| h_n \right|^2x}$, then the differential of $f(x)$ can be obtained by
\begin{equation} \label{eq:daoshu_1}
	f^{\prime}\left( x \right) =\frac{-\left( \sqrt{x\left| h_m \right|^2+1}-1 \right) ^2}{2x^2\gamma \left| h_m \right|^2\sqrt{x\gamma \left| h_m \right|^2+1}}\le 0 \,\left( x\ge 0 \right).
\end{equation}

So, (\ref{eq:tuilun_1}) can be further written as
\begin{equation} \label{eq:jielun_1}
	\mathrm{R}_{m}^{\left( m,n \right)}\ge \mathrm{R}_{m}^{OMA}
	\Leftarrow \epsilon _n\le \frac{\sqrt{1+\,\gamma \,\left| h_m \right|^2T_N}-1}{\gamma \,\left| h_m \right|^2 T_N}.
\end{equation}

Then, considering the fairness constraint for User-$n$, $\epsilon_n$ should satisfy
\begin{equation} \label{eq:tuilun_2}
	\begin{aligned}
		&\mathrm{R}_{n}^{\left( m,n \right)}\ge \mathrm{R}_{n, \rm Nyquist}^{OMA} \,\,\Leftarrow \,\, \mathrm{R}_{n, \rm Nyquist}^{\left( m,n \right)}\ge \mathrm{R}_{n, \rm Nyquist}^{OMA} 
		\\
		&\Leftrightarrow {\log _2\left( \frac{1+\epsilon _n\gamma \left| h_n \right|^2 T_N}{\sqrt{1+\gamma \left| h_n \right| T_N}} \right)} \ge 0
		\\
		&\Leftrightarrow \frac{1+\epsilon _n\gamma \left| h_n \right|^2 T_N}{\sqrt{1+\gamma \left| h_n \right| T_N}}\ge 1
		\Leftrightarrow \epsilon _n\ge \frac{\sqrt{1+\,\gamma \,\left| h_n \right|^2T_N}-1}{\,\gamma \,\left| h_n \right|^2T_N}\,\, .
	\end{aligned}
\end{equation}


\begin{figure*}[hb!]
	\hrulefill
	\begin{equation} \label{eq:differential_sc} \tag{13}
		\frac{\mathrm{d}\left( \mathrm{R}_{m}^{\left( m,n \right)}+\mathrm{R}_{n}^{\left( m,n \right)} \right)}{\mathrm{d}\epsilon _n}=-\frac{1}{\pi \alpha (1+\beta)}\int_0^{\pi}{\frac{G\left( \omega \right) \,\gamma \,\left( \left| h_m \right|^2-\left| h_n \right|^2 \right)}{\ln 2 \,\left( G\left( \omega \right) \,\varepsilon _n\,\gamma \,\left| h_m \right|^2+1 \right) \,\left( G\left( \omega \right) \,\varepsilon _n\,\gamma \,\left| h_n \right|^2+1 \right)}\mathrm{d}\omega}>0
	\end{equation}
\end{figure*}

To further determine the value of $\epsilon_n$, we present the differential of ASR as (\ref{eq:differential_sc}). Hence, the upper bound of $\epsilon_n$ should be employed to maximize the ASR of the whole system. The value of $\epsilon_n$ can be determined as
\setcounter {equation} {13}
\begin{equation} \label{eq:epsilon_sc} 
	\epsilon _n=\frac{\sqrt{1+\,\gamma \,\left| h_m \right|^2T_N}-1}{\gamma \,\left| h_m \right|^2 T_N}.
\end{equation}

Finally, we verify the range of $\epsilon_n$ ($0\le \epsilon_n \le 1$). According to (\ref{eq:tuilun_1})-(\ref{eq:tuilun_2}) and L'Hospital's rule \cite{stewart2015calculus}, the range of $\epsilon_n$ can be obtained as
\begin{equation}
	\begin{aligned}
		0 &\le \frac{\sqrt{1+\,\gamma \,\left| h_n \right|^2T_N}-1}{\gamma \,\left| h_n \right|^2 T_N} \le \epsilon_n \\ 
		& = \frac{\sqrt{1+\,\gamma \,\left| h_m \right|^2T_N}-1}{\gamma \,\left| h_m \right|^2 T_N} \le 0.5  < 1.
	\end{aligned}
\end{equation}
So, the value of $\epsilon_n$ in \eqref{eq:epsilon_sc} is proved to satisfy its original assumption. 

\begin{figure*}[b!]
\begin{equation} \label{eq:diff_1} \tag{18}
	\mathrm{\bar{R}}_3-\mathrm{\bar{R}}_2
	\\
	=\frac{1}{\pi \alpha (1+\beta)}\int_0^{\pi}{\log _2\left[ 1+\frac{\left( \left| h_4 \right|^2-\left| h_3 \right|^2 \right) \left( \bar{\epsilon}_1-\bar{\epsilon}_2 \right) \gamma G\left( \omega \right)}{\left( 1+\bar{\epsilon}_1\gamma \left| h_3 \right|^2G\left( \omega \right) \right) \left( 1+\bar{\epsilon}_2\gamma \left| h_4 \right|^2G\left( \omega \right) \right)} \right]}{\rm d}\omega 
	\\
	\ge 0
\end{equation}
\end{figure*}

\begin{figure*} [b!]
\hrulefill
\begin{equation} \tag{19}
	\begin{aligned} \label{eq:diff_2}
		&\mathrm{\bar{R}}_2-\mathrm{\bar{R}}_1
		\\
		=&\frac{1}{\pi \alpha (1+\beta)}\int_0^{\pi}{\log _2\left[ \frac{\left( 1+\bar{\epsilon}_1\gamma \left| h_3 \right|^2G\left( \omega \right) \right) \left( 1+\bar{\epsilon}_2\gamma \left| h_4 \right|^2G\left( \omega \right) \right) \left( 1+\frac{\left| h_2 \right|^2\left( 1-\bar{\epsilon}_2 \right) G\left( \omega \right)}{\bar{\epsilon}_2\left| h_2 \right|^2G\left( \omega \right) +\gamma ^{-1}} \right)}{\left( 1+\bar{\epsilon}_1\gamma \left| h_2 \right|^2G\left( \omega \right) \right) \left( 1+\bar{\epsilon}_3\gamma \left| h_4 \right|^2G\left( \omega \right) \right) \left( 1+\frac{\left| h_3 \right|^2\left( 1-\bar{\epsilon}_3 \right) G\left( \omega \right)}{\bar{\epsilon}_3\left| h_3 \right|^2G\left( \omega \right) +\gamma ^{-1}} \right)} \right]}\rm{d}\omega \\
		=&\frac{1}{\pi \alpha (1+\beta)}\int_0^{\pi}{\log _2\left[ \frac{\left( 1+\bar{\epsilon}_1\gamma \left| h_3 \right|^2G\left( \omega \right) \right) \left( 1+\bar{\epsilon}_2\gamma \left| h_4 \right|^2G\left( \omega \right) \right) \left( \left| h_2 \right|^2G\left( \omega \right) +\gamma ^{-1} \right) \left( \bar{\epsilon}_3\left| h_3 \right|^2G\left( \omega \right) +\gamma ^{-1} \right)}{\left( 1+\bar{\epsilon}_1\gamma \left| h_2 \right|^2G\left( \omega \right) \right) \left( 1+\bar{\epsilon}_3\gamma \left| h_4 \right|^2G\left( \omega \right) \right) \left( \left| h_3 \right|^2G\left( \omega \right) +\gamma ^{-1} \right) \left( \bar{\epsilon}_2\left| h_2 \right|^2G\left( \omega \right) +\gamma ^{-1} \right)} \right]}\rm{d}\omega 
		\\
		\ge &\frac{1}{\pi \alpha (1+\beta)}\int_0^{\pi}{\log _2\left[ \frac{\left( 1+\bar{\epsilon}_1\gamma \left| h_3 \right|^2G\left( \omega \right) \right) \left( 1+\bar{\epsilon}_2\gamma \left| h_3 \right|^2G\left( \omega \right) \right) \left( \left| h_2 \right|^2G\left( \omega \right) +\gamma ^{-1} \right) \left( \bar{\epsilon}_3\left| h_3 \right|^2G\left( \omega \right) +\gamma ^{-1} \right)}{\left( 1+\bar{\epsilon}_1\gamma \left| h_2 \right|^2G\left( \omega \right) \right) \left( 1+\bar{\epsilon}_3\gamma \left| h_2 \right|^2G\left( \omega \right) \right) \left( \left| h_3 \right|^2G\left( \omega \right) +\gamma ^{-1} \right) \left( \bar{\epsilon}_2\left| h_2 \right|^2G\left( \omega \right) +\gamma ^{-1} \right)} \right]}\rm{d}\omega 
		\\
		=&\frac{1}{\pi \alpha (1+\beta)}\int_0^{\pi}{\left[ \log _2\left( \frac{1+\bar{\epsilon}_1\gamma \left| h_3 \right|^2G\left( \omega \right)}{\left| h_3 \right|^2G\left( \omega \right) +\gamma ^{-1}} \right) -\log _2\left( \frac{1+\bar{\epsilon}_1\gamma \left| h_2 \right|^2G\left( \omega \right)}{\left| h_2 \right|^2G\left( \omega \right) +\gamma ^{-1}} \right) \right]}\rm{d}\omega 
		\\
		\ge & 0
	\end{aligned}
\end{equation}
\end{figure*}

\subsection{User Pairing and Power Allocation in 4-user FTN-based SC-NOMA}
Here, we consider an FTN-based SC-NOMA system with 4 users that will be divided into 2 pairs. The power allocation in each pair obeys \eqref{eq:epsilon_sc}. So, there are 3 possible user pairing schemes defined as follows.

\begin{itemize}
\item Scheme-1: \{User-1, User-2\}, \{User-3, User-4\}
\item Scheme-2: \{User-1, User-3\}, \{User-2, User-4\}
\item Scheme-3: \{User-1, User-4\}, \{User-2, User-3\}
\end{itemize}

And $\epsilon_n$ for the second user in each scheme can be obtained as
\begin{equation}
\left\{ \,\,\, \begin{aligned}
	\epsilon _{2}^{\left( 1,2 \right)}&=\epsilon _{3}^{\left( 1,3 \right)}=\epsilon _{4}^{\left( 1,4 \right)}=\frac{\sqrt{1+\,\gamma \,\left| h_1 \right|^2T_N}-1}{\gamma \,\left| h_1 \right|T_N}\triangleq \bar{\epsilon}_1\\
	\epsilon _{3}^{\left( 2,3 \right)}&=\epsilon _{4}^{\left( 2,4 \right)}=\frac{\sqrt{1+\,\gamma \,\left| h_2 \right|^2T_N}-1}{\gamma \,\left| h_2 \right|T_N}\triangleq \bar{\epsilon}_2\\
	\epsilon _{4}^{\left( 3,4 \right)}&=\frac{\sqrt{1+\,\gamma \,\left| h_3 \right|^2T_N}-1}{\gamma \,\left| h_3 \right|T_N}\triangleq \bar{\epsilon}_3\\
\end{aligned} \right. ,
\\
\end{equation}
where $\epsilon _{m}^{\left( m,n \right)}$ and $\epsilon _{n}^{\left( m,n \right)}$ represent the power allocation coefficients for User-$m$ and User-$n$ in the $(m,n)$-pair, and $\bar{\epsilon}_3\le \bar{\epsilon}_2\le \bar{\epsilon}_1$. According to \eqref{eq:sc_rate_near}, the relationship among available rates for the first user can be obtained as
\begin{equation}
	\left\{ \,\,\, \begin{aligned}
		\mathrm{R}_{1}^{\left( 1,4 \right)}&=\mathrm{R}_{1}^{\left( 1,3 \right)}=\mathrm{R}_{1}^{\left( 1,2 \right)}\\
		\mathrm{R}_{2}^{\left( 2,3 \right)}&=\mathrm{R}_{2}^{\left( 2,4 \right)}\\
	\end{aligned} \right. .
	\\
\end{equation}

Then, we compare ASRs of different user pairing schemes by calculating their differences. $\bar R_1$, $\bar R_2$ and $\bar R_3$ represent the ASRs of 3 schemes, respectively. According to the differences of ASRs between 3 schemes, as demonstrated in \eqref{eq:diff_1} and \eqref{eq:diff_2}, the conclusion can be drawn as
\setcounter {equation} {19}
\begin{equation} \label{eq:rate_relationship}
	\bar {\mathrm{R}}_3 \ge \bar{\mathrm{R}}_2 \ge \bar {\mathrm{R}}_1.
\end{equation}

Finally, Scheme-3 is proved to be the optimal user pairing scheme in the 4-user FTN-based SC-NOMA considering the power allocation in \eqref{eq:epsilon_sc}. 

\subsection{User Pairing and Power Allocation in $2K$-user FTN-based SC-NOMA}
In this subsection, we employ \textit{reductio ad absurdum} to obtain the optimal user pairing for FTN-based SC-NOMA with $2K$ users considering the power allocation in \eqref{eq:epsilon_sc}.

\emph{Theorem 1:} Considering the power allocation algorithm in \eqref{eq:epsilon_sc}, in FTN-based SC-NOMA with $2K$ users, the optimal user pairing scheme is that each User-$k$ ($k=1,2\cdots2K$) pairs with User-($2K+1-k$).

\emph{Proof:} For any user pairing scheme named Scheme-A that doesn't satisfy \emph{Theorem 1}, we assume that User-$\hat k$ ($1\le\hat k\le K$) is the first user that doesn't pair with User-($2K+1-\hat k$). Or rather, for any $l\le \hat k$, User-$l$ pairs with User-($2K+1-l$).

Then, we use User-$a$ and User-$b$ to denote the users paired with User-$\hat k$ and User-($2K+1-\hat k$), respectively. The scheme can be written as
\begin{itemize}
	\item Scheme-A: \\..., \{User-$\hat k$, User-$a$\},\{User-$b$, User-($2K+1-\hat k$)\}, ...\\
	$\left(\hat k<a<(2K+1-\hat k) , \,\,\, \hat k<b<(2K+1-\hat k)\right)$
\end{itemize}

A comparison scheme name Scheme-B is constructed as follows. 

\begin{itemize}
	\item Scheme-B \\..., \{User-$\hat k$, User-$(2K+1-\hat k)$\},\{User-$a$, User-$b$\}, ...
\end{itemize}

Here, we use $\bar R_A$ and $\bar R_B$ to represent the ASRs of Scheme-A and Scheme-B. According to \eqref{eq:rate_relationship}, the relationship of $\bar R_A$ and $\bar R_B$ can be demonstrated as
\begin{equation}
	\bar R_A \le \bar R_B.
\end{equation}

So, for any user pairing scheme that doesn't satisfy \emph{Theorem 1}, there always exists at least one different scheme that can achieve a higher ASR than it. Hereto, user pairing in \emph{Theorem 1} is proved to be the optimal scheme for $2K$-user FTN-based SC-NOMA with the power allocation in \eqref{eq:epsilon_sc}.

\section{System Model of the FTN-based MIMO-NOMA}  \label{sec:mimo_model}
This section considers the FTN-based downlink MIMO-NOMA with $2K$ users to be paired in each resource (time/frequency/code) block. The number of antennas in the BS and each user terminal (UT) are $K$ and $N$, respectively. The users' distribution is the same as that in Section \ref{sec:sc_model}. Also, as assumed in FTN-based SC-NOMA, both the BS and UTs know exactly the CSI of all the users.

Herein, we use $\tilde{\boldsymbol{s}}_i\in \mathbb{C}^{K\times 1}$ to denote the $i$-th transmitted symbols on BS's antennas in the time domain, which contains symbols of all the $K$ pairs. The symbol for the $n$-th pair in $\tilde{\boldsymbol{s}}_i$ can be written as

\begin{equation}
\tilde{s}_{i,n}=\sqrt{\epsilon _{1,n}} \hat s_{i,1,n}+\sqrt{\epsilon _{2,n}} \hat s_{i,2,n} ,
\end{equation}
where $\epsilon_{1,n}$ and $\epsilon_{2,n}$ is the power allocation coefficients for the first and second user in the $n$-th user pair ($n=1,2,\cdots, K$) and $\epsilon_{1, n}+\epsilon_{2, n}=1$. $\hat s_{i,j,n}$ is the $i$-th signal of the $j$-th user in the $n$-th pair. Also, the first user and second user are considered as the weak and strong user, respectively.

Then, the transmitted symbols will be precoded as
\begin{equation}
	\mathbf{x}_i=\mathbf{W\tilde{s}}_i=\left[ \mathbf{w}_1,\mathbf{w}_2,...,\mathbf{w}_N \right] \tilde{\mathbf{s}}_i ,
\end{equation}
where $\mathbf{W}\in\mathbb{C}^{K\times K}$ is the precoding matrix. Finally, the $i$-th received symbol for the $j$-th user in the $n$-th pair $\mathbf{r}_{i,j,n}\in \mathbb{C}^{N\times 1}$ can be written as
\begin{equation}
	\begin{aligned}
		\mathbf{r}_{i,j,n}&=\sum_{k=-\infty}^{+\infty}{\mathbf{H}_{j,n}\mathbf{x}_{i-k}}g(\left( i-k \right) \alpha T_N)+\mathbf{\tilde{n}}_{j,n}\left( i\alpha T_N \right)\\
		&=\mathbf{H}_{j,n}\mathbf{w}_n\tilde s_{i,n}g(i\alpha T_N)+\sum_{k=1,k\ne i}^N{\mathbf{H}_{n,i}\mathbf{w}_k \tilde s_{i,k}g\left( i\alpha T_N \right)}\\
		&+\sum_{k=-\infty ,k\ne 0}^{+\infty}{\mathbf{H}_{j,n}\mathbf{W\tilde{s}}_k}g\left( \left( i-k \right) \alpha T_N \right) +\mathbf{\tilde{n}}_{j,n}\left( i\alpha T_N \right) ,
	\end{aligned}
\end{equation}
where $\mathbf{H}_{j,n}=\tilde{\mathbf{H}}_{j,n}/\sqrt{(d_{j,n})^3}\in \mathbb{C}^{N\times K}$ is the channel gain matrix of the $j$-th user in the $n$-th pair. $d_{j,n}$ represents the distance from the BS to the $j$-th user in the $n$-th pair. And $\tilde{\mathbf{H}}_{j,n}$ obeys the Rayleigh distribution. $\tilde{\mathbf n}_{j,n}(t)\in \mathbb C^{N\times 1}$ is the colored noise for $j$-th user in the $n$-th pair which is obtained by the convolution of white noise and the matched filter. As present in Section \ref{sec:sc_model}, PSD of the white noise is $N_0$.

In the receiver, the received symbols in different antennas should be combined to improve the signal-to-noise ratio (SNR). And the $i$-th post-proceeded symbol for $j$-th user in the $n$-th pair can be written as
\begin{equation}
	\begin{aligned}
		y_{i,j,n}&=\mathbf{v}_{j,n}^{H}\mathbf{r}_{i,j,n}\\
		& =\mathbf{v}_{j,n}^{H}\mathbf{H}_{j,n}\mathbf{W}\sum_{k=-\infty}^{+\infty}{\mathbf{s}_k}g\left((i-k)\alpha T_N\right) \\
		& +\mathbf{v}_{j,n}^{H}\mathbf{\tilde{n}}_{i,j,n}\left( i\alpha T_N \right),
	\end{aligned}
\end{equation}
where $\mathbf{v}_{j,n}\in\mathbb{C}^{N\times 1}$ is the combining vector for the $j$-th user in the $n$-th pair.

\begin{figure*}[b!]
\hrulefill
\begin{equation} \label{eq:mimo_sinr1}  \tag{27}
	\mathrm{SINR}_{1,n}\left( \omega \right) =\frac{\left( 1-\epsilon _{2,n} \right) \left| \mathbf{v}_{1,n}^{H}\mathbf{H}_{1,n}\mathbf{w}_n \right|^2 G\left(\omega \right)}{\left. \left\| \mathbf{v}_{1,n}^{H} \right\| \right. ^2\gamma ^{-1}+\left( \epsilon _{1,n}\left| \mathbf{v}_{1,n}^{H}\mathbf{H}_{1,n}\mathbf{w}_n \right|^2+\sum_{k=1,k\ne n}^N{\left| \mathbf{v}_{1,n}^{H}\mathbf{H}_{1,n}\mathbf{w}_k \right|^2} \right) G\left(\omega \right)}
\end{equation}
\end{figure*}

\begin{figure*} [b!]
\begin{equation} \label{eq:mimo_rate1}  \tag{35}
	R_{1,n} =\frac{1}{\pi \alpha (1+\beta)} \int_0^\pi \log_2 \left(1+\frac{\left( 1-\epsilon _{2,n} \right) \left| \mathbf{v}_{1,n}^{H}\mathbf{H}_{1,n}\mathbf{w}_n \right|^2 G\left( \omega \right)}{\left. \left\| \mathbf{v}_{1,n}^{H} \right\| \right. ^2\gamma ^{-1}+\epsilon _{2,n}\left| \mathbf{v}_{1,n}^{H}\mathbf{H}_{1,n}\mathbf{w}_n \right|^2G\left( \omega \right)} \right) {\rm d} \omega
\end{equation}
\end{figure*}

Similar to Section \ref{sec:sc_model}, in the MIMO-NOMA system, the second user will firstly decode and remove the signal for the first user and then decode its own symbols. The SINR for the second user can be written as
\begin{equation} \label{eq:mimo_sinr2}
\begin{aligned}
	&\mathrm{SINR}_{2,n}\left( \omega \right) \\
	& =\frac{\epsilon _{2,n}\left| \mathbf{v}_{2,n}^{H}\mathbf{H}_{2,n}\mathbf{w}_n \right|^2 G\left( \omega \right)}{\left\| \left. \mathbf{v}_{2,n}^{H} \right. \right\|_2 ^2\gamma ^{-1}+\sum_{k=1,k\ne n}^N{\left| \mathbf{v}_{2,n}^{H}\mathbf{H}_{2,n}\mathbf{w}_k \right|^2}G\left(\omega \right) \cdot}
\end{aligned}
\end{equation}

The first user will directly decode its own symbols, while the signal for the second user is treated as interference. The SINR of the first user can be written as \eqref{eq:mimo_sinr1}.

\section{User Pairing and Power Allocation for FTN-based MIMO-NOMA System}  \label{sec:up_mimo_noma}
\subsection{Interference Cancellation}
Although the SINRs in \eqref{eq:mimo_sinr2} and \eqref{eq:mimo_sinr1} contain inter-pair interference and seem to be complicated, with adequately constructed precoding and combining vectors, the interference can be effectively eliminated. Firstly, we employ zero-forcing (ZF) precoding and equal gain combining (EGC) for the second user in each pair. The precoding vector for the $n$-th pair can be obtained by solving
\setcounter{equation}{27}
\begin{equation} \label{eq:cal_precoding}
	\left[\mathbf{g}_{2,1}^{H}, \ldots, \mathbf{g}_{2, n-1}^{H}, \mathbf{g}_{2, n+1}^{H}, \ldots, \mathbf{g}_{2, K}^{H}\right]^{H} \mathbf{w}_{n}=\mathbf{0},
\end{equation}
where $\mathbf{g}_{2, n} \in \mathbb{C}^{1 \times K}$ is the equivalent channel vector which can be written as
\begin{equation}
	\mathbf{g}_{2, n}=\mathbf{v}_{2, n}^{H} \mathbf{H}_{2, n}.
\end{equation}

And the combining vector $\mathbf{v}_{2,n}$ should be constructed as
\begin{equation}
	\mathbf{v}_{2, n}=\frac{1}{\sqrt{N}}[1,1, \ldots, 1]^{T}.
\end{equation}

Finally, the rate of the second user in the $n$-th pair can be present as
\begin{equation} \label{eq:mimo_rate2}
\begin{aligned}
	&R_{2,n} \\=&\frac{1}{\pi \alpha (1+\beta)}
	\int_0^\pi \log_2\left(1+ \epsilon _{2,n}\gamma\left| \mathbf{v}_{2,n}^{H}\mathbf{H}_{2,n}\mathbf{w}_n \right|^2 G\left(\omega \right)\right) {\rm d}\omega.
\end{aligned}
\end{equation}

Then, we construct the combining vector for the first user in each pair as
\begin{equation} \label{eq:cal_combining}
	{\mathbf{v}_{1,n}=\mathbf{H}_{1,n}\mathbf{u}_{1,n}},
\end{equation}
where $\mathbf{u}_{1,n}$ can be obtained by solving
\begin{equation}
	{\mathbf{u}_{1,n}^{H}\left[ \mathbf{\hat g}_{1,1},...,\mathbf{\hat g}_{1,n-1},\mathbf{\hat g}_{1,n+1},...,\mathbf{\hat g}_{1,K} \right] =0},
\end{equation}
where $\mathbf{\hat g}_{1,n} \in \mathbb{C} ^{K\times 1}$ is defined as
\begin{equation}
	{\mathbf{\hat g}_{1,n}=\mathbf{H}_{1,n}^{H}\mathbf{H}_{1,n}\mathbf{w}_n}.
\end{equation}

The rate of the first user in the $n$-th pair can be written as \eqref{eq:mimo_rate1}. 

Finally, considering the multiplexing loss, the rates of the first user and second user in the MIMO-OMA are present as
\setcounter{equation}{35}
\begin{equation}
	\begin{aligned}
		\mathrm{R}_{1,n}^{\mathrm{OMA}} =& \frac{1}{2\pi\alpha (1+\beta)} \cdot \\
		 &\int_0^\pi \log _2\left( 1+\frac{\gamma \left| \mathbf{v}_{1,n}^{H}\mathbf{H}_{1,n}\mathbf{w}_n \right|^2 G(\omega)}{\left\| \mathbf{v}_{1,n}^{H} \right\| ^2} \right) {\rm d} \omega
	\end{aligned}
\end{equation}
and
\begin{equation}
	\begin{aligned}
		\mathrm{R}_{2,n}^{\mathrm{OMA}} =&\frac{1}{2\pi\alpha(1+\beta)} \cdot \\
	& \int_0^\pi\log _2\left( 1+\gamma \left| \mathbf{v}_{2,n}^{H}\mathbf{H}_{2,n}\mathbf{w}_n \right|^2 G(\omega)\right) {\rm d} \omega.
	\end{aligned}
\end{equation}

\subsection{Power Allocation Considering User Fairness}
As seen, the format of \eqref{eq:mimo_rate2} and \eqref{eq:mimo_rate1} is quite similar to \eqref{eq:sc_rate_near} and \eqref{eq:sc_rate_far}. So, it is easy to consider that a similar derivation as Section \ref{sub:up_2_user} can be employed to obtain the power allocation scheme for the FTN-based MIMO-NOMA.

Here, we define
\begin{equation} \label{eq:mimo_h_m}
	\tilde {h}_m = \frac{ \mathbf{v}_{1,n}^{H}\mathbf{H}_{1,n}\mathbf{w}_n }{\left\| \mathbf{v}_{1,n}^{H} \right\|}
\end{equation}
and
\begin{equation} \label{eq:mimo_h_n}
	\tilde{h}_n =  \mathbf{v}_{2,n}^{H}\mathbf{H}_{2,n}\mathbf{w}_n
\end{equation}
to replace $h_m$ and $h_n$ in Section \ref{sec:up_sc_noma}, respectively. Then, according to \eqref{eq:epsilon_sc}, with the same derivation and verification in Section \ref{sub:up_2_user}, the power allocation coefficient for FTN-based MIMO-NOMA can be obtained as
\begin{equation}
	\begin{aligned}
	\epsilon _n&=\frac{\sqrt{1+\,\gamma \,\left| \tilde{h}_m \right|^2T_N}-1}{\gamma \,\left| \tilde{h}_m \right|^2 T_N} \\
	&= \frac{\left\| \mathbf{v}_{1,n}^{H} \right\|\left(\sqrt{\left\| \mathbf{v}_{1,n}^{H} \right\|^2+\,\gamma \,\left| \mathbf{v}_{1,n}^{H}\mathbf{H}_{1,n}\mathbf{w}_n \right|^2T_N}-\left\| \mathbf{v}_{1,n}^{H} \right\|\right)}{\gamma \,\left| \mathbf{v}_{1,n}^{H}\mathbf{H}_{1,n}\mathbf{w}_n \right|^2 T_N}.
	\end{aligned}
\end{equation}

\subsection{Dynamic User Pairing}
The effective channel gains shown in \eqref{eq:mimo_h_m} and \eqref{eq:mimo_h_n}  contain precoding and combining vectors that can not be determined at the beginning. Hence, the corresponding user pairing is a dynamic process accompanied by the construction of precoding and combining vectors. Here, we design a dynamic user pairing scheme for the FTN-based MIMO-NOMA, as shown in Algorithm \ref{algo:up}, where $\mathring{\mathbf H}$ and $\mathring{\mathbf H_j}$ represents the Gram-Schmidt orthogonalization operated on matrix $\mathbf H$ and $\mathbf H_j$. And $\|\cdot\|_F$ denotes the Frobenius norm of the matrix.

\begin{algorithm}
	\caption{Dynamic User Pairing for FTN-based MIMO-NOMA System} \label{algo:up}
	\KwIn {$\mathbf{H}_1, \mathbf{H}_2, \cdots \mathbf{H}_{2K}$ $\left(\left\| \mathbf{H}_1 \right\|_F^2 \ge \left\| \mathbf{H}_2 \right\|_F^2 \ge \cdots \left\| \mathbf{H}_{2K} \right\|_F^2\right)$}
	\KwOut {$\mathbf{U}_1, \mathbf{U}_2$, $\mathbf W$, $\mathbf V$;}
	Initialize $\mathbf{U}_1=\mathbf{U}_2=\varnothing$, $\mathbf W = \mathbf V = \varnothing$, $\mathbf G_1=\{1,2,\cdots K\}$, $\mathbf G_2=\{K+1, K+2, \cdots 2K\}$; \\
	\For
	{$i\rightarrow 1:K$}
	{
	\eIf{i=1}
	{$k = {\rm{argmax}} \left\{\left\| \mathbf H_j \right\|_F^2, \, j\in\mathbf G_1 \right\}$; \\
	}
	{
	Chordal distance \cite{ko2012multiuser}: $k  = {\rm{argmax}}$ \\ \, \\
	$\left\{ \left\| \mathbf{H}_j \right\| _{F}^{0.1}\cdot \left\| \mathbf{\mathring{H}\mathring{H}}^H-\mathbf{\mathring{H}}_j{\mathbf{\mathring{H}}_j}^H \right\| _F, j\in \mathbf{G}_1 \right\}$;
	}
	
	${\mathbf H} = \left[{\mathbf H}, \mathbf H_k \right] $, \, $\mathbf U_1=\mathbf U_1\cup k$, \, $\mathbf G_1 = \mathbf G_1\setminus k$;
	}
	Calculate $\mathbf W=\left[ \mathbf w_1, \mathbf w_2, \cdots \mathbf w_K \right]$ with \eqref{eq:cal_precoding};
	
	
	\For{$i\rightarrow 1:K$}
	{
	Calculate possible combining vectors $\tilde{\mathbf v}_j$ with \eqref{eq:cal_combining} for all $j\in \mathbf G_2$, \\
	$\tilde u = \mathbf u(i)$, $k = {\rm{argmin}}\left\{\left| \tilde{\mathbf v}_j^H \mathbf H_{\tilde u} \mathbf w_i \right|^2, j\in \mathbf G_2\right\}$;  \\
	$\mathbf U_2 = \mathbf U_2 \cup k$, $\mathbf V = \left[\mathbf V, \tilde{\mathbf v}_k\right]$, $\mathbf G_2 = \mathbf G_2 \setminus k$;
	}
\end{algorithm}


\section{Numerical Result}  \label{sec:simulation}
This section presents comprehensive simulations on the proposed user pairing and power allocation for both FTN-based SC-NOMA and MIMO-NOMA systems. For the convenience of comparison, all the simulated scenarios employ the same power allocation scheme presented in \eqref{eq:epsilon_sc}. The curve labeled random UP means the NOMA with random user pairing, while the other curves without specific labels represent NOMA with the proposed user pairing scheme. The channel gain is assumed to obey standard Rayleigh distribution at the position of 100m away from the BS. In the MIMO system, the BS and UTs both equips 8 antennas (i.e., $K=8$ and $N=8$). And the radius $d$ is set as 500m. $\lambda$ and $\mu$ are 0.8 and 0.2, respectively.


\subsection{ASR of the FTN-based SC-NOMA versus SNR} \label{sub:asr_sc}
\begin{figure} [ht]
	\centering
	\includegraphics[width=0.84 \linewidth]{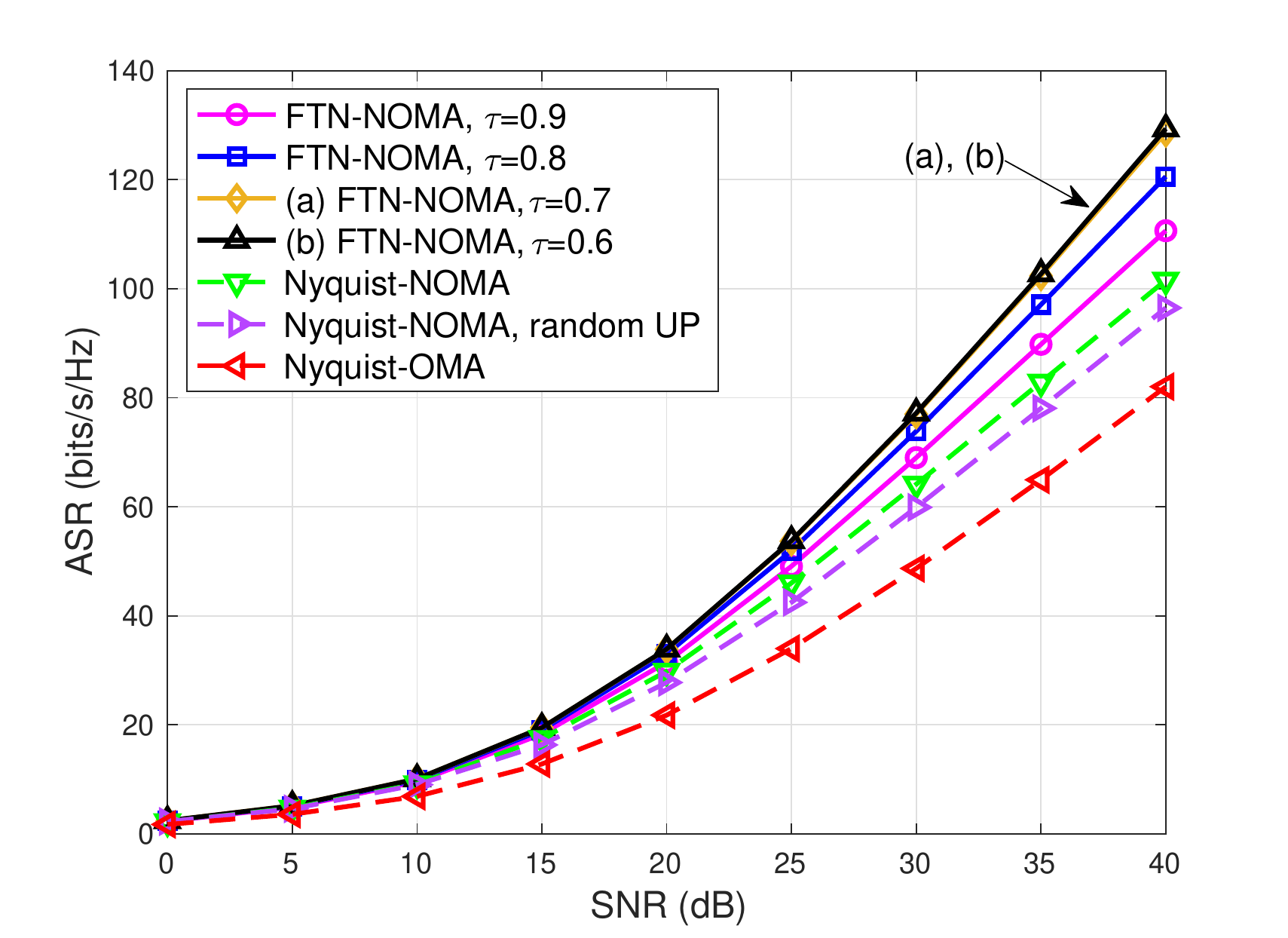}
	\caption{ASRs in different SC scenarios with 32 users where $\beta=0.5$}
	\label{fig:asr_sc_0p5}
\end{figure}

\begin{figure} [ht]
	\centering
	\includegraphics[width=0.84 \linewidth]{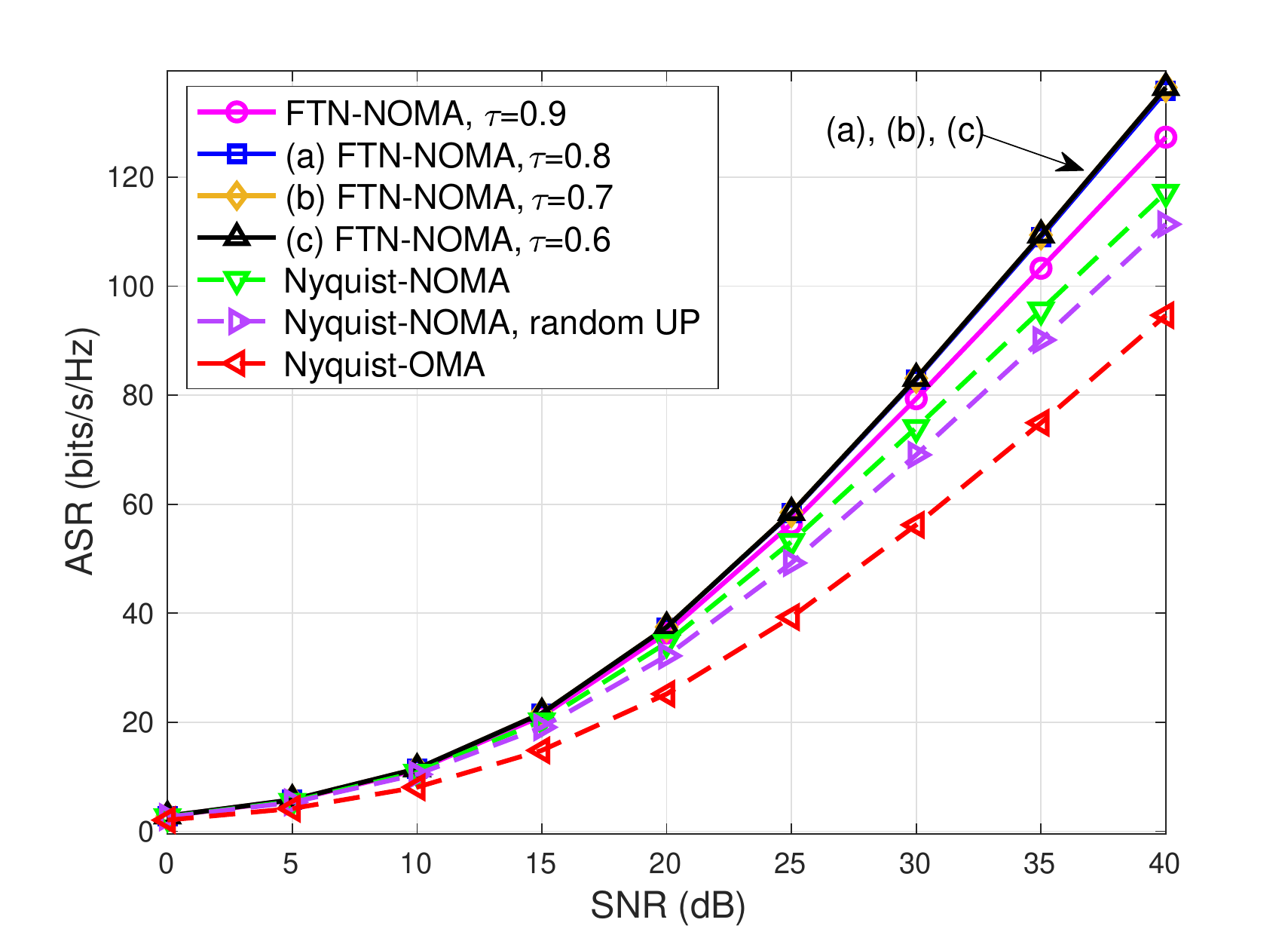}
	\caption{ASRs in different SC scenarios with 32 users where $\beta=0.3$}
	\label{fig:asr_sc_0p3}
\end{figure}

Fig. \ref{fig:asr_sc_0p5} and Fig. \ref{fig:asr_sc_0p3} demonstrate the comparison among ASRs in different SC scenarios with 32 users where $\beta=0.5$ and $\beta=0.3$, respectively. As seen, the FTN-based SC-NOMA schemes achieve obviously better performances than the Nyquist-NOMA and Nyquist-OMA with ASR gains of up to 30\% and 58\% where SNR=40dB with $\beta=0.5$, respectively. It shows the power of FTN-based SC-NOMA with the proposed user pairing and power allocation to improve the system ASR performance. Also, the effectiveness of the proposed user pairing is proved by its performance gain over the random user pairing scheme.

\begin{figure*}[h!]
	\centering
	\makebox[\textwidth][c]{\includegraphics[width=1.05\textwidth]{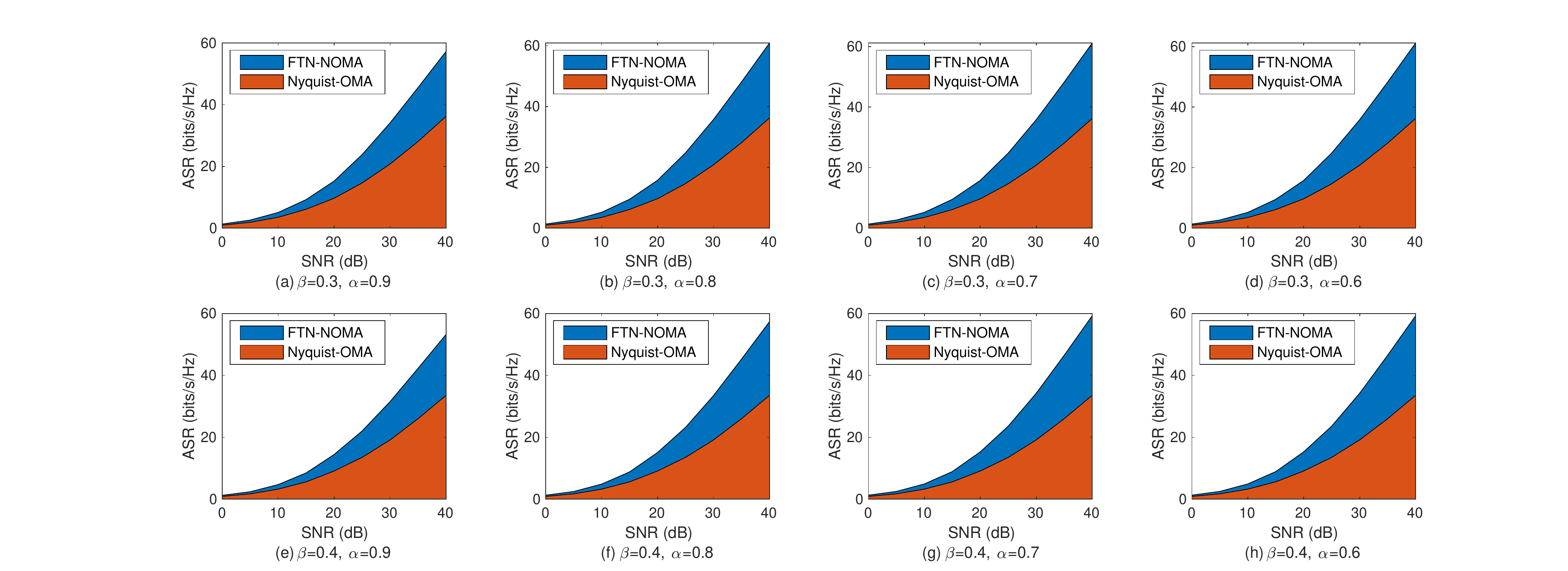}}
	\caption{ASR gains of the FTN-based MIMO-NOMA over the conventional Nyquist-criterion MIMO-OMA with 16 users}
	\label{fig:asr_all}
\end{figure*}

It is worth noting that some curves overlap in Fig. \ref{fig:asr_sc_0p5} and Fig.\ref{fig:asr_sc_0p3}. This phenomenon is consistent with the conclusion in \cite{rusek2009constrained} that capacity of FTN exceeds that of Nyquist signals only when $\alpha>1/(1+\beta)$. Also, when bandwidth is limited, with the increase of SNR, the capapcity of FTN signaling with different $\beta$ and $\alpha<=1/(1+\beta)$ values will approach the same ultimate capacity with the given noise PDF.

\subsection{ASR of the FTN-based SC-NOMA versus Number of Users}
\begin{figure}[ht]
	\centering
	\includegraphics[width=0.84 \linewidth]{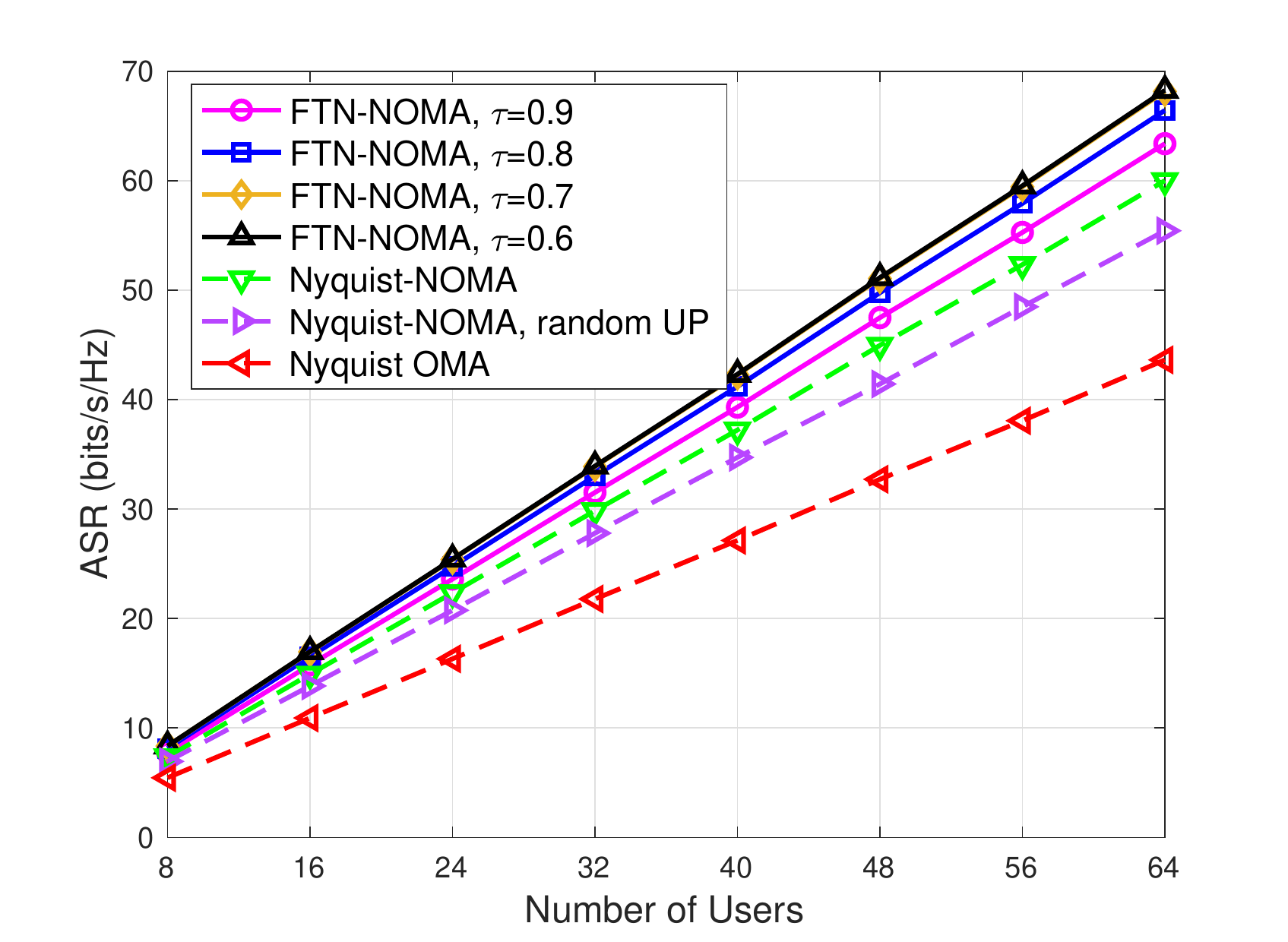}
	\caption{ASRs in different SC scenarios with varying number of users where $\beta=0.5$ and SNR=20dB}
	\label{fig:userk}
\end{figure}


Fig. \ref{fig:userk} shows the ASR of different SC scenarios with a varying number of users, where $\beta=0.5$ and SNR=20dB. As shown, similar to Section \ref{sub:asr_sc}, the ASRs of the FTN-based SC-NOMA with the proposed user pairing and power allocation outperform those of all the other scenarios. The ASR gains of the proposed scheme over Nyquist-NOMA and Nyquist-OMA achieve up to 15\% and 60\%, respectively. And more importantly, with the increase in the number of users, the ASRs grow successively without any performance floors. It verifies the effectiveness of the FTN-based SC-NOMA and the proposed scheme in communications with massive users.

\subsection{QoS of the FTN-based SC-NOMA (User Fairness)}
\begin{figure}[ht]
	\centering
	\includegraphics[width=0.84 \linewidth]{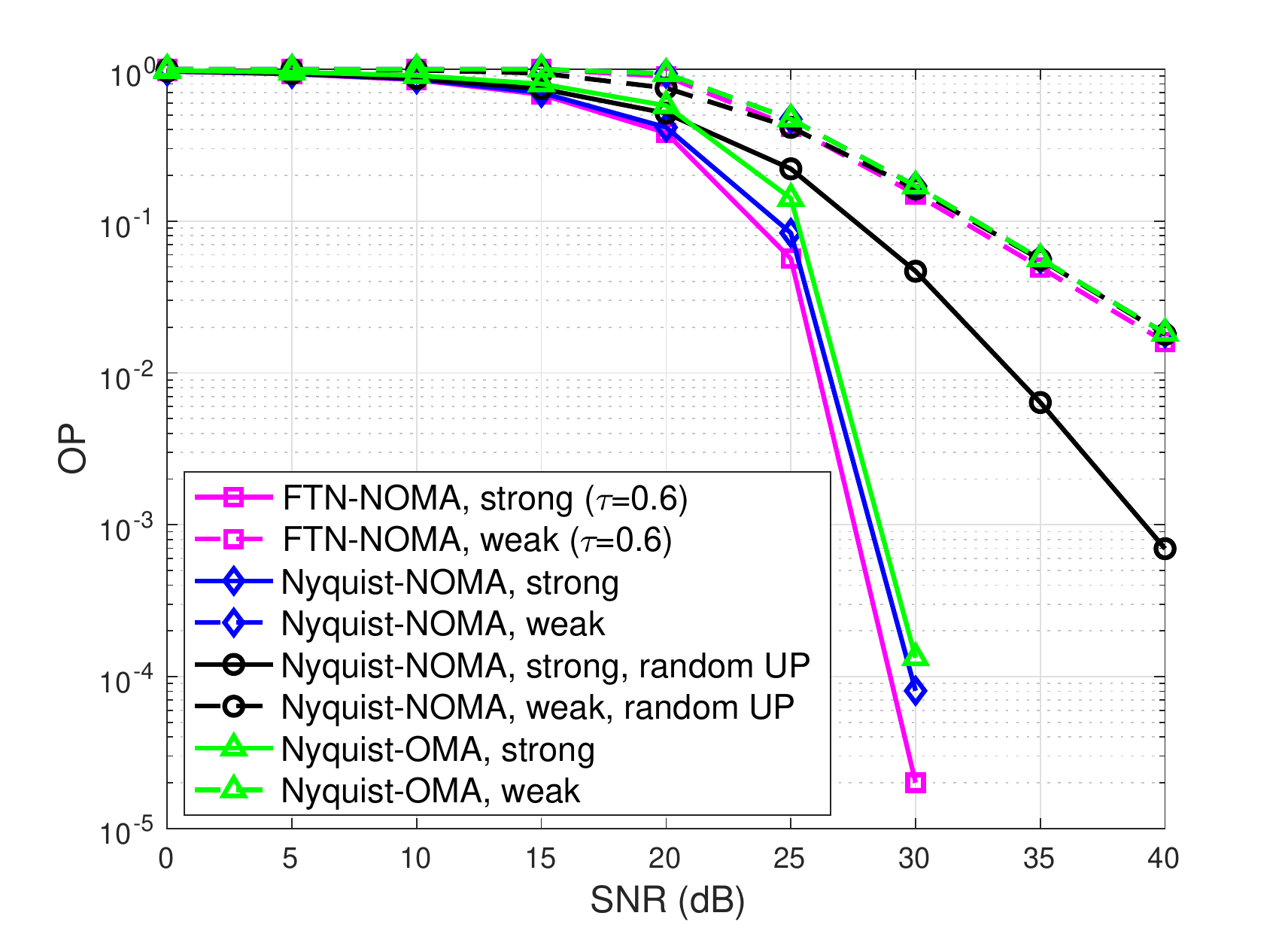}
	\caption{OPs in different SC scenarios with 32 users where $\beta=0.5$}
	\label{fig:op_sc}
\end{figure}

Apart from ASR, the communications should also meet the requirement of QoS to guarantee the necessary rate for each user to keep reliable transmission. Here, we consider a fixed QoS requirement, that is, the rate must be greater than a fixed threshold, or the user will suffer from the outage. Fig. \ref{fig:op_sc} demonstrates the outage probability (OP) for different SC scenarios where the rate thresholds for the strong and weak users are set as 1 bit/s/Hz and 0.5 bit/s/Hz, respectively. 

As shown, the FTN-based SC-NOMA with the proposed user pairing and power allocation can effectively improve the reliability of the communications by its lower OPs over the conventional Nyquist-OMA. Also, with similar or even better performance for weak users, the proposed user pairing shows significantly advantages in strong users' OPs over random user pairing (e.g., a gain of 12dB when considering OP=$10^{-3}$), which means the proposed user pairing can better utilize the strong users' channel conditions and achieve a higher QoS.

\subsection{ASR of the FTN-based MIMO-NOMA versus SNR}
\begin{figure} [ht]
	\centering
	\includegraphics[width=0.84 \linewidth]{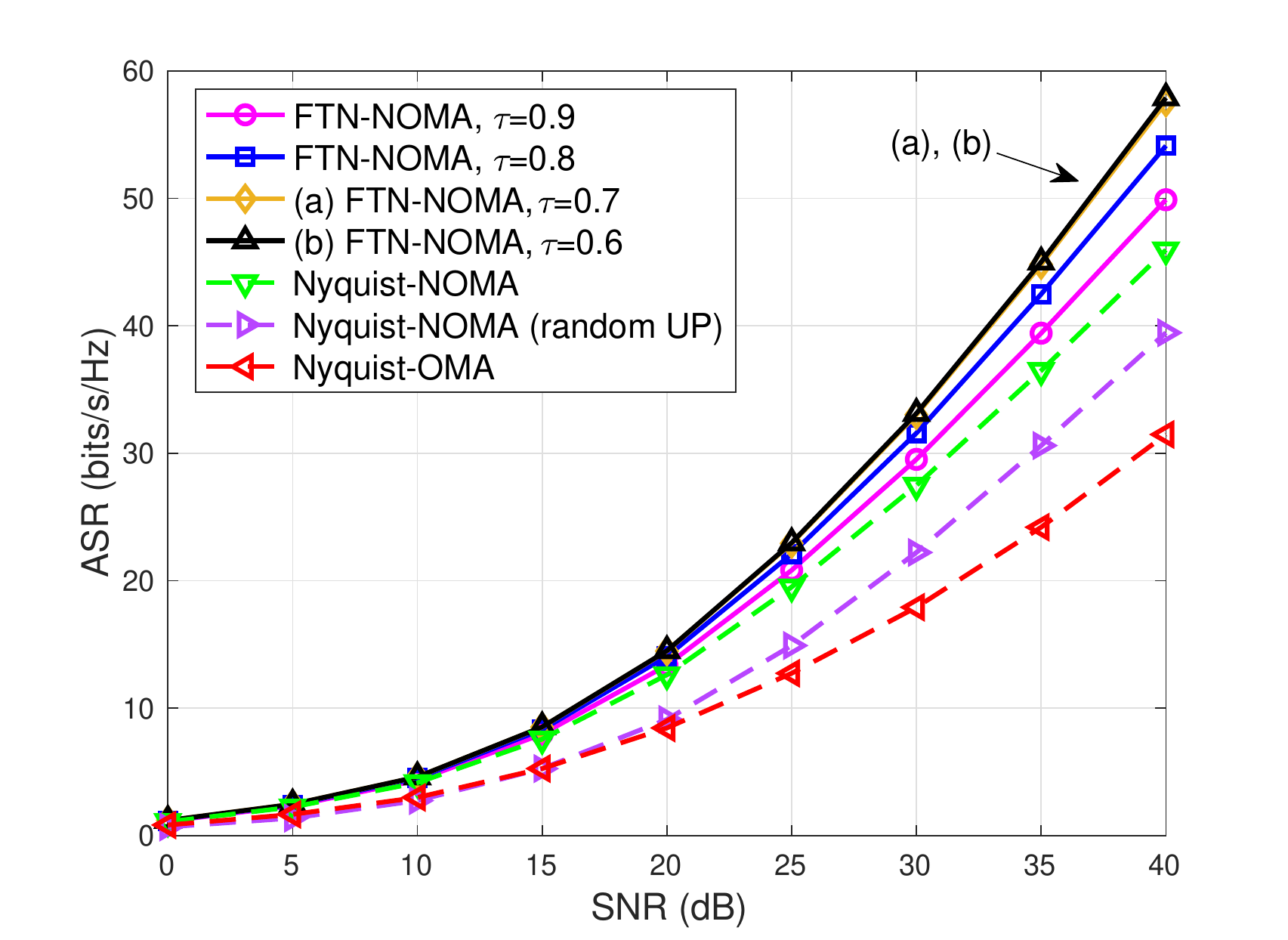}
	\caption{ASRs in different MIMO scenarios with 16 users}
	\label{fig:asr_mimo}
\end{figure}
Fig. \ref{fig:asr_mimo} demonstrates the ASRs of different MIMO scenarios with 16 users. Apart from the curve labeled $\beta=0.3$, all the other scenarios apply the roll-off factor $\beta = 0.5$. Similar to Section \ref{sub:asr_sc}, the FTN-based MIMO-NOMA with the proposed user pairing achieves the best performance, which proves the effectiveness of the proposed scheme. Its performance gain over Nyquis-NOMA and Nyquist-OMA systems can achieve up to 25\% and 85\%, respectively, in high SNRs.

It is worth noting that all the users in Fig. \ref{fig:asr_mimo} share the same resource block. While in the SC-NOMA system, only 2 users within one pair can occupy the same resource. Hence, the ASR in Fig. \ref{fig:asr_mimo} corresponds to the sum rate of each pair in the SC-NOMA system. Benefiting from the multiplexing gain provided by multiple antennas, each resource block in the MIMO-NOMA can carry more users and business than that in the SC-NOMA.

Finally, Fig. \ref{fig:asr_all} comprehensively demonstrates the ASR gains of the FTN-based MIMO-NOMA with the proposed scheme over the conventional Nyquist-OMA transmission.


\subsection{QoS of the FTN-based MIMO-NOMA (User Fairness)}
\begin{figure}[ht]
	\centering
	\includegraphics[width=0.84 \linewidth]{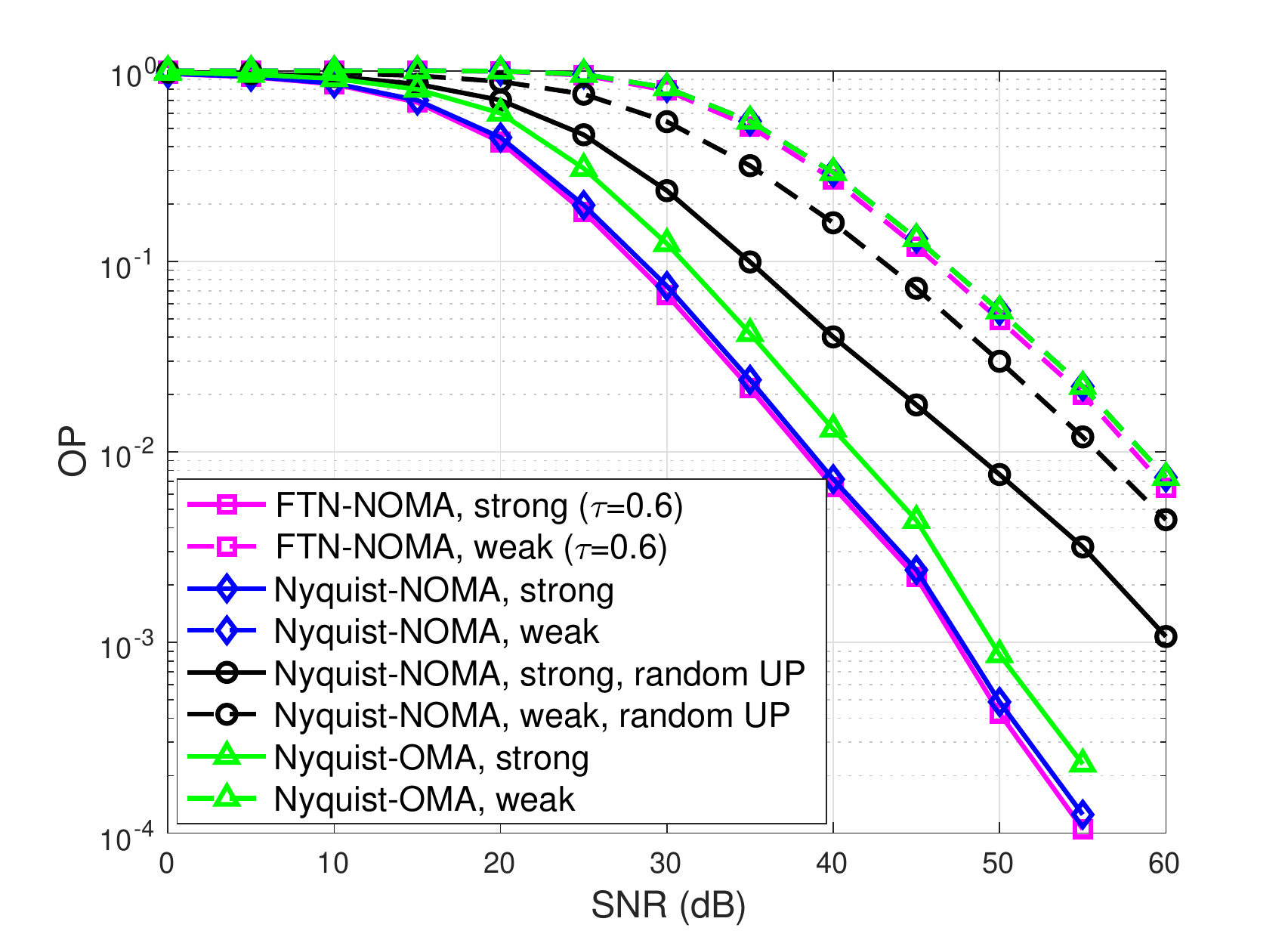}
	\caption{OPs in different MIMO scenarios with 16 users where $\beta=0.5$}
	\label{fig:op_mimo}
\end{figure}
Fig. \ref{fig:op_mimo} demonstrates the OPs of different MIMO scenarios with 16 users where $\beta=0.5$. The fixed QoS requirement is considered and the rate thresholds for the strong and weak users are set as 1 bit/s/Hz and 0.5 bit/s/Hz, respectively. As shown, with similar OP for weak users, the proposed user pairing and power allocation can effectively reduce the OP of strong users compared with random user pairing. And the proposed scheme also shows great advantages over the conventional MIMO-OMA system.

\section{Conclusion}  \label{sec:conclusion}
In this paper, we solve the problem of user pairing and power allocation for FTN-based NOMA systems considering user fairness. And this paper is, as far as we know, the first paper studying on this issue. We first derive an effective power allocation for the FTN-based SC-NOMA with 2 users, considering user fairness. Then, it is generalized to 4-user and 2K-user scenarios while the optimal user pairing scheme is proposed. Afterward, we obtain the power allocation for the FTN-based MIMO-NOMA with a similar derivation and propose a dynamic user pairing scheme. Finally, we carry out comprehensive simulations on our proposed schemes. And the results have shown the advantage of the proposed user pairing and power allocation in both ASR and QoS performance.

Although effective schemes have been proposed to make the FTN-based NOMA technology available, we think there are still some challenges on this issue. For example, in FTN-based MIMO-NOMA, how to pre-allocate users for different resource blocks to further improve ASR for all users? When the number of receiving antennas is less than that of transmitting ones, how to design a  user pairing scheme? These issues will be studied in our future works.



\ifCLASSOPTIONcaptionsoff
  \newpage
\fi



\bibliographystyle{IEEEtran}
\bibliography{database.bib}
%

%

%
%
%




\end{document}